\documentclass[11pt]{article}
\usepackage{amsmath}
\begin{document}
\title{\bf Gauge equivalent universes in 5d Kaluza-Klein theory}
\author{Gyeong Yun Jun, Pyung Seong Kwon
\footnote{E-mail:bskwon@star.ks.ac.kr}\\
{\small Department of Physics, Kyungsung University, Pusan
608-736, Korea}}
\date{}
\maketitle \thispagestyle{empty} \baselineskip 8.5mm
\begin{abstract}
 We examine in the framework of 5d Kaluza-Klein theory the gauge
 equivalence of $x^5$-dependent cosmological solutions each of
which describes
 in the 4d sector an arbitrarily evolving isotropic, homogeneous
universe with some pure
 gauge. We find that (1)within a certain time scale $\tau_c$
 (which is characterized by the compactification radius $R_c$) any
 arbitrarily evolving 4d universe
 is allowed to exist by field equations, and these 4d universes with
appropriate
  pure gauges are all gauge equivalent as long as they are of the
same topology. (2)Outside $\tau_c$ the gauge equivalence
disappears and the
 evolution of the universe is fixed by field equations.
\end{abstract}
\medskip
\begin{center}
\end{center}
\newpage
\baselineskip 6.0mm

\setcounter{page}{1} Caused by brane world scenarios\cite{111} the
5d cosmology\footnote{For the traditional 5d cosmological
solutions, see ref.\cite{222,555}} has been severely modified
during the past several years\cite{333}. The general feature of
the new scenarios is that the scale of extra dimensions is not so
small as expected by the traditional Kaluza-Klein theories, and
the dependence of metric components on these extra dimensions
becomes important. In this paper we will examine a 5d cosmology
which is conventional in the sense that the theory possesses a
$U(1)$ gauge symmetry at the massless level\footnote{Thus, in this
paper, the extra dimension is an analogue of an internal
Calabi-Yau space, rather than a $S^1 / Z_2$ orbifold}, but where
metric components have $x^5$-dependence.

Recently, there has been an argument\cite{444} in 5d Kaluza-Klein
theory that an evolving universe may be related with a static
universe by a gauge transformation. The authors have used the
simplest $x^5$-dependent cosmological solution to the vacuum
Einstein equation to show that the time degree of freedom of an
evolving universe can be absorbed by a gauge transformation into
the fifth dimension, and consequently the evolving universe turns
into a static one. The solution considered in ref.\cite{444} took
the form of the Tolman metric
\begin{equation}
ds^2=-dt^2 + R^2(t,x^5)d\Omega^2_k + e^{{\mu}(t,x^5)} (dx^5)^2
\end{equation}
with
\begin{equation}
R^2(t,x^5)=R^2_0 + \alpha^2f_0 [x^5 + (t-t_0)/\alpha]^2,
\end{equation}
and $e^{\mu(t,x^5)}$ being related with $R^2(t,x^5)$ by the
equation \footnote{In this paper we will use the same notations as
those in ref.\cite{444}; i.e., the ``prime'' denotes the
$x^5$-derivative, while ``overdot''the time-derivative etc.}
\begin{equation}
e^{\mu} = \frac{R'^2}{\alpha^2}
        =  \frac{\alpha^2 f_0^2[x^5 +(t-t_0)/\alpha]^2}{R_0^2
+\alpha^2f_0[x^5
       +(t-t_0)/\alpha]^2},
\end{equation}
where $d\Omega^2_k$ is the metric of the 3d volume with constant
curvature $ k =1, 0, -1$, and $R_0, f_0$ and $\alpha$ are all
(integral) constants, and in particular
\begin{equation}
\alpha =\pm (k+f_0)^{1/2}.
\end{equation}
The solution (1) reduces to a static solution of the form
\begin{equation}
ds^2 = -dt^2 + \tilde{R}^2(\tilde{x}^5) d\Omega^2_k +
e^{\tilde{\mu}(\tilde{x}^5)} [d\tilde{x}^5 + \kappa
\tilde{A_0}(t)dt ]^2
\end{equation}
once we perform a gauge transformation
\begin{equation}
x^5 \to \tilde{x}^5 =x^5 +(t-t_0)/\alpha ,
\end{equation}
which shows that the time degree of freedom of an evolving
universe can be gauged away by a U(1) gauge transformation, and in
compensation for this a pure gauge comes into being. But here we
should notice that the fifth coordinates $x^5$ and $\tilde{x}^5$
are both compact variables whose principal values can not exceed
$\pm 2\pi R_c$ (where $R_c$ represents the compactification radius
of the fifth dimension), while $t$ is a noncompact variable which
can be arbitrarily large. This means that the relation in eq.(6)
fails to hold once $|(t-t_0)/\alpha |$ exceeds the value $2\pi
R_c$, and the solutions in eqs.(1) and (5) are not gauge
equivalent anymore in the region $(t-t_0)/\alpha
> 2\pi R_c$. This is remarkable because it indicates the
possibility that any two different 4d universes may be entirely
equivalent at the early stage of evolution within a certain time
scale characterized by $R_c$.

Let us consider a metric
\begin{equation}
ds^2=-N^2(t,x^5)dt^2 + R^2(t,x^5)d\Omega^2_k +e^{\mu(t,x^5)}[dx^5
+N^5(t)dt]^2,
\end{equation}
where $N(t,x^5)$ is the lapse function which has been introduced
for the time being, and $N^5(t)$ is the fifth (and the only
non-vanishing) component of the shift vector $N^A(t)$. With
$N(t,x^5)=1$ and $N^5(t)=0$ the metric (7) reduces to the Tolman
metric. The action on the other hand is given by the
Hilbert-Einstein action
\begin{eqnarray}
I_g & = & -\frac{1}{2\kappa} \int d^{5}x
\sqrt{-\mathop{g}\limits^5}\mathop{\boldsymbol{R}}\limits^{5} \\
    & = & -\frac{1}{2\kappa} \int dt d^4x
    N\sqrt{\mathop{g}\limits^4}(\mathop{\boldsymbol{R}}\limits^4 +K_
{\mu
    \nu}K^{\mu \nu}-K^2) + surface\hspace{2mm} terms \\
   & \equiv & \int dt d^4x \mathcal{L}_g + surface \hspace{2mm}
terms
\end{eqnarray}
plus perhaps some matter action
\begin{equation}
I_m=\int d^5x \mathcal{L}_m
\end{equation}
which has not been given in a definite form. In eq.(9),
$\mathop{\boldsymbol{R}}\limits^4$ and $K_{\mu \nu}$ are the Ricci
scalar and the second fundamental form of the 4d spacelike
hypersurface, and they are given by
\begin{equation}
\mathop{\boldsymbol{R}}\limits^4 =
\frac{6k}{R^2}-e^{-\mu}[6\frac{R^{\prime
\prime}}{R}+6(\frac{R^{\prime}}{R})^2 -3\mu^{\prime}
\frac{R^{\prime}}{R} ],
\end{equation}
\begin{equation}
K_{\mu \nu} =\frac{1}{2N}[\partial_0{\mathop{g}\limits^4}_{\mu
\nu}-({\mathop{\nabla}\limits^4}_{\mu} N_{\nu}
+{\mathop{\nabla}\limits^4}_{\nu} N_{\mu})]
    \\,
\hspace{0.4cm}(\mu, \nu = 1,2,3,5;\hspace{2pt} \partial_0 \equiv
\partial_t).
\end{equation}
where ${\mathop{\nabla}\limits^4}_{\mu}$ denotes the covariant
derivative associated with ${\mathop{g}\limits^4}_{\mu \nu}$ which
is induced on the hypersurface. The action in (9) can be put into
a canonical form by introducing canonical momenta
$\boldsymbol{\pi}^{\mu \nu}$ which are defined by
\begin{equation}
\boldsymbol{\pi}^{\mu \nu} \equiv \frac{\partial
\mathcal{L}_g}{\partial(\partial_0 {\mathop{g}\limits^4}_{\mu
\nu})} = -\frac{\sqrt{ \mathop{g}\limits^4}}{2\kappa}(K^{\mu
\nu}-{\mathop{g}\limits^4}{}^{\mu \nu}K) .
\end{equation}
For a given metric (7) the non-vanishing components of
$\boldsymbol{\pi}^{\mu \nu}$ are
\begin{equation}
\boldsymbol{\pi}^{ij}= \frac{\sqrt{\mathop{g}\limits^4}}{2\kappa}[
\frac{2}{N}(\frac{1}{R}D_0 R)+\frac{1}{2N}(D_0
\mu)]{\mathop{g}\limits^4}{}^{ij}, \hspace{0.4cm}(i, j = 1,2,3),
\end{equation}
\begin{equation}
\boldsymbol{\pi}^{55}=\frac{\sqrt{\mathop{g}\limits^4}}{2\kappa}
[\frac{3}{N}(\frac{1}{R}D_0 R)]e^{-\mu} ,
\end{equation}
where the derivative $D_0$ is defined by $D_0 \equiv \partial_0 -
N^5 \partial_5$. With these $\boldsymbol{\pi}^{\mu \nu}$ the
Lagrangian density $\mathcal{L}_g$ is then written as
\begin{equation}
\mathcal{L}_g = \boldsymbol{\pi}^{\mu \nu} \partial _0
{\mathop{g}\limits^4}_{\mu \nu} -(N \mathcal{H} +N_5
\mathcal{H}^5),
\end{equation}
where
\begin{equation}
\mathcal{H} \equiv -
\frac{2\kappa}{\sqrt{\mathop{g}\limits^4}}(\boldsymbol{\pi}_{\mu
\nu}\boldsymbol{\pi}^{\mu \nu}- \frac{1}{3}\boldsymbol{\pi}^2)+
\frac{\sqrt{\mathop{g}\limits^4}}{2\kappa}
\mathop{\boldsymbol{R}}\limits^4
\end{equation}
and
\begin{equation}
\mathcal{H}^5 \equiv
-2{\mathop{\nabla}\limits^4}_{\nu}\boldsymbol{\pi}^{5 \nu}
\end{equation}
are related with the Hamiltonian $H$ by the equation
\begin{equation}
H= \int d^4x ( N\mathcal{H} +N_5 \mathcal{H}^5).
\end{equation}
 In the presence of matter fields (i.e. for $\mathcal{L}_m \neq 0$)
the field equations obtained by varying $N$ and $N_5$ are the
Hamiltonian constraint
\begin{equation}
\mathcal{H} =N^2\sqrt{\mathop{g}\limits^4}T^{00} ,
\end{equation}
and the momentum constraint
\begin{equation}
\mathcal{H}^5 =-N\sqrt{\mathop{g}\limits^4}(T^{05} + N^5T^{00}),
\end{equation}
where $T^{AB}$ are expectation values of the stress-energy tensor
of 5d matter fields.

Having found constraint equations we now set $N \equiv 1$ and $N^5
\equiv \kappa A_0$; namely, we are considering a metric
\begin{equation}
ds^2=-dt^2 + R^2(t,x^5)d\Omega^2_k +e^{\mu(t, x^5)}[dx^5 + \kappa
A_0(t) dt]^2.
\end{equation}
Upon this setting the Hamiltonian constraint in eq.(18) can be
recast into more suggestive form:
\begin{equation}
\mathcal{H} = \frac{\sqrt{\mathop{g}\limits^4}}{\kappa}G_{00}
+\sqrt{\mathop{g}\limits^4}\rho_A ,
\end{equation}
where
\begin{equation}
G_{00}=3(\frac{\dot{R}}{R})^2 +\frac{3}{2}
\dot{\mu}\frac{\dot{R}}{R} +\frac{1}{2}
\mathop{\boldsymbol{R}}\limits^4
\end{equation}
is the 00-component of the 5d Einstein tensor derived from the
Tolman metric (1), and
\begin{equation}
\rho_A \equiv \frac{1}{2\kappa}[(\kappa A_0)H_1 +(\kappa A_0)^2H_2
]
\end{equation}
with
\begin{equation}
H_1 \equiv -12(\frac{\dot{R}}{R})(\frac{R^{\prime}}{R})
-3\frac{\dot{R}}{R}\mu^{\prime}-3\frac{R^{\prime}}{R}\dot{\mu} ,
\end{equation}
\begin{equation}
H_2 \equiv 6(\frac{R^{\prime}}{R})^2
+3\frac{R^{\prime}}{R}\mu^{\prime}
\end{equation}
is the energy density associated with the pure gauge $A_0$. This
is quite surprising. Being a physically non-observable quantity a
pure gauge essentially does not contribute to the energy (or
Lagrangian) of spacetime due to the vanishing of the field
strength. Indeed a pure gauge does not play any role in ordinary
4d theories of spacetime. This, however, is not true anymore in
the 5d theory under discussion. Eq.(24) shows that part of the
energy is engaged in the dynamics of spacetime, but the rest is
stored in $\rho_A$ in the form of a pure gauge. $\rho_A$ manifests
itself once the metric components have $x^5$-dependence.

 To find the solution to the constraint equations (21) and (22),
 $T^{AB}$ must be definitely given.  In our discussion we will
focus our attention on the case $T^{AB}=0$ because it is not only
simple, but it is of particular interest in the context of the
discussion in ref.\cite{444}. As for the solution to field
equations we consider an ansatz of the form
\begin{equation}
R^2(t,x^5)=R_0^2+\alpha^2f_0[x^5 + \kappa \xi(t)]^2,
\end{equation}
\begin{equation}
e^{\mu(t, x^5)}  =  \frac{{R^{\prime}}^2}{\alpha^2}=\frac{\alpha^2
f_0^2[x^5 + \kappa \xi(t)]^2}{R^2_0 +\alpha^2 f_0[x^5 + \kappa
\xi(t)]^2} \hspace{3mm},
\end{equation}
which is obviously a generalization of eqs.(2)
 and (3), and where both $R^2$ and $e^{\mu}$ are expressed in
terms of a single function $\xi(t)$, meaning that the dynamics of
the universe is entirely described by $\xi(t)$ alone. The function
$\xi(t)$ is of course to be determined by field equations for a
given (in our case, zero) matter distribution. However, $\xi(t)$
is subject to the gauge transformation; $\xi(t)$ experiences a
transformation
\begin{equation}
\xi (t) \to \tilde{\xi}(t) = \xi(t) -\Lambda(t)
\end{equation}
under the gauge transformation
\begin{equation}
x^5 \to \tilde{x}^5 =x^5 + \kappa \Lambda(t) ,
\end{equation}
\begin{equation}
A_0(t) \to \tilde{A}_0(t) = A_0(t)-\dot{\Lambda}(t).
\end{equation}
 Since the gauge parameter $\Lambda(t)$ is totally
arbitrary eq.(31) implies that the gauge transformation relates
two arbitrarily different universes with dynamics described,
respectively, by $\tilde{\xi}(t)$ and $\xi(t)$. In the 4d sector
these correspond to two arbitrarily evolving 4d isotropic
universes with scales described by two arbitrary functions $R(t,
x^5)$ and $\tilde{R}(t, \tilde{x}^5)$. Indeed, using (24) and (25)
one can convert the Hamiltonian constraint (21) into the Einstein
equation for a homogeneous, isotropic 4d cosmology
\begin{equation}
 3(\frac{\dot{R}}{R})^2 + \frac{3k}{R^2} = \kappa
 \mathop{\rho}\limits^4  \!_{\rm eff} \,\,\,,
\end{equation}
where the 4d effective source $\mathop{\rho}\limits^4 \!_{\rm
eff}$ is given by\footnote{Note that the relation $e^\mu = {R
^\prime}^2 / \alpha^2$ has been used to obtain eq.(35). We also
have set $T^{00} = 0$ ; thus, all the  gauge equivalent 5d
universes considered in this paper have the same(i.e., zero) 5d
energy density.}
\begin{equation}
\mathop{\rho}\limits^4  \!_{\rm eff}= - \frac{3}{2\kappa} \dot\mu
\frac{\dot{R}}{R} + \frac{3}{\kappa}\frac{\alpha^2}{R^2} - \rho_A
\,\,\,.
\end{equation}
The first term in eq.(35) is a conventional term, typical of
ordinary 5d cosmology\cite{555}, while the remaining two terms are
new terms that appear only when the metric components have
$x^5$-dependence. Now it is important to note that
$\mathop{\rho}\limits^4 \!_{\rm eff}$ contains the energy density
$\rho_A$, whose value essentially depends on $A_0$. Since $A_0$ is
subject to the gauge transformation (33) this implies that
$\mathop{\rho}\limits^4 \!_{\rm eff}$ can have arbitrary values
depending on $\Lambda(t)$. Thus, any two 4d universes with
arbitrarily different 4d sources can be related by the gauge
transformation (32) and (33). As discussed before the gauge
transformation (32) is valid only within a certain time scale
characterized by $R_c$; to be precise, for the gauge
transformation (32) the time scale is given by $|\kappa
\Lambda(t)|\sim \pi R_c$ (we will call this scale $\tau_c$).
Outside $\tau_c$, the transformation (32), and consequently the
relation (31) are broken down and the gauge equivalence between
two universes disappears. Further, in the region sufficiently far
from $\tau_c$ (i.e. for $\kappa \xi \gg \pi R_c$) it is convenient
to introduce a new time variable $\tilde{t}$ which is defined by
\begin{equation}
x^5 +\kappa \xi(t) \equiv \kappa \xi (\tilde{t}).
\end{equation}
In that region $\tilde{t}$ approximates $t$, and
$x^5$-dependencies of the metric components can be neglected.

Turning back to field equations one can show that $\mathcal{H}^5$
in (19) identically vanishes upon substituting (29) and (30):
\begin{equation}
\mathcal{H}^5 =0 ,
\end{equation}
that is, the momentum constraint is automatically satisfied by the
given ansatz. However, $\mathcal{H}$ and $\rho_A$ in eqs.(18) and
(26) do not vanish identically upon substituting (29) and (30);
they are calculated to give
\begin{equation}
\mathcal{H}=\frac{3
f_0}{\kappa}\frac{\sqrt{\mathop{g}\limits^4}}{R^2}[\kappa^2\alpha^2
(A_0-\dot{\xi})^2-1],
\end{equation}
\begin{equation}
\sqrt{\mathop{g}\limits^4}\rho_A = 3\kappa \alpha^2 f_0
\frac{\sqrt{\mathop{g}\limits^4}}{R^2}[A_0(A_0-2\dot{\xi})].
\end{equation}
  In eq.(39), $\rho_A$ vanishes not only for
$A_0=0$, it also vanishes when $A_0 =2\dot{\xi}$. Also with
$A_0=0$ the solution to the Hamiltonian constraint $\mathcal{H}=0$
coincides with the solution described by (2) and (3) as it should
be. However, it is important to note that the equation $
\mathcal{H} =0$ does not generally determine $A_0$ and $\xi$
separately; it only determines the combination $A_0- \dot{\xi}$,
which is gauge invariant under the combined transformations (31)
and (33). In fact, within $\tau_c$, $A_0$ and $\xi$ are not
determined even by field equations as can be seen in the
followings. The Lagrangian density calculated from eq.(9) takes
the form
\begin{equation}
\mathcal{L}_g=\frac{3f_0}{\kappa}\frac{\sqrt{\mathop{g}\limits^4}}
{R^2}[\kappa^2\alpha^2(A_0-\dot{\xi})^2+1],
\end{equation}
but one can verify that the above $\mathcal{L}_g$ is simply a sum
of total derivative terms\footnote{One can use the relation
$e^{\mu} = {R^{\prime}}^2/\alpha^2$ in eq.(30) to show that
$\sqrt{\mathop{g}\limits^4}/R^2=\partial_5
[\sqrt{\mathop{g}\limits^4}R^{\prime} e^{-\mu}/2 \alpha^2 R$].}.
Thus the variation of the action $I_g$ always vanishes for any
$\xi$ and $A_0$:
\begin{equation}
 \delta_{\xi} I_g =\delta_{A_0} I_g = 0,
 \end{equation}
which means that the field equations are trivially satisfied by
any $\xi$ and $A_0$, and any universe described by ($\xi(t)$,
$A_0(t)$) is allowed within $\tau_c$. This supports our conjecture
that the gauge equivalent universes are equally allowed in this
region.

What about outside $\tau_c$ then? Since $x^5$-dependencies of
$R^2$ and $e^{\mu}$ disappear in this region (see eq.(36)) all the
covariant derivatives $D_0$ in $\boldsymbol{\pi}^{\mu \nu}$ (or
$K^{\mu \nu}$) are replaced by ordinary derivatives $\partial_0$,
and therefore the Lagrangian (or the action) does not include
$A_0$ anymore. In this case the action in fact takes the same form
as the conventional action associated with the standard 5d
Robertson-Walker metric, so the solution that minimize the action
is expected to be\cite{444,555}
\begin{equation}
ds^2=-dt^2 + [R^2_0 - k (t-t_0)^2 ]d\Omega^2_k +
\frac{(t-t_0)^2}{[R^2_0 - k (t-t_0)^2 ]}[dx^5 + \kappa A_0(t)
dt]^2,
\end{equation}
where $A_0(t)$ is arbitrary\footnote{$A_0(t)$ is arbitrary because
it does not appear in the action, and the equation of motion for
$A_0(t)$ does not exist.}. From this we see that the pure gauge
can exist even in the region outside $\tau_c$. However, in this
region the pure gauge does not play any physically important role
(actually, it is insensible) because it manifests itself always
through the covariant derivative $D_0 =
\partial_0 - \kappa A_0 \partial_5$, but the metric components are
$x^5$-independent there. In fact, $dx^5 + \kappa A_0(t) dt$ in
(42) can always be replaced simply by $dx^5$ by an appropriate
coordinate transformation, then we recover the ordinary 5d
Robertson-Walker metric representing a radiation-dominated
universe.

The above discussion may be extended to the general case. For
instance,  the ansatz in (29) and (30) may not be relevant to the
case of $T^{AB}\neq 0$. Recall that it is just a generalization of
the solution to vacuum($T^{AB}=0$) field equations. The most
general ansatz for $R^2$ and $e^{\mu}$ would be in fact of the
form
\begin{equation}
R^2(t,x^5)=R^2(x^5 +\kappa \xi(t)) ,
\end{equation}
\begin{equation}
e^{\mu(t, x^5)} = \frac{{R^{\prime}}^2}{\alpha^2},
\end{equation}
and one can verify that the momentum constraint $\mathcal{H}^5$
identically vanishes as before upon substituting (43) and (44):
\begin{equation}
\mathcal{H}^5 =0.
\end{equation}
This equation agrees with (22) when
\begin{equation}
N^5=\kappa A_0 =-T^{05}/T^{00} ,
\end{equation}
which suggests that the pure gauge $A_0$ can be interpreted (in a
5d sense) as a momentum density (normalized by $\kappa T^{00}$)
along the fifth direction. The Hamiltonian constraint on the other
hand takes the form
\begin{equation}
\mathcal{H} =\frac{3}{\kappa}
\frac{\sqrt{\mathop{g}\limits^4}}{R^2}[
\frac{\kappa^2}{2}(A_0-\dot{\xi})^2(R^2)^{\prime \prime}-f_0],
\end{equation}
which is obviously a generalization of eq.(38), and still the
constraint equation (21) does not determine $A_0$ and $\xi$
separately\footnote{In the case of $T^{00} \neq 0$, the quantity
$\sqrt{\mathop{g}\limits^4} T^{00}$ must be taken to be gauge
invariant in order for $\mathcal{H}$ to be gauge invariant} .
Finally, The Lagrangian density $\mathcal{L}_g$ (i.e., the
generalization of (40)) is calculated to give
\begin{equation}
\mathcal{L}_g =\frac{3}{\kappa}
\frac{\sqrt{\mathop{g}\limits^4}}{R^2}[\frac{\kappa^2}{2}(A_0-\dot
{\xi})^2(R^2)^{\prime \prime}+f_0],
\end{equation}
and one can verify that the terms in eq.(48) only contribute to
surface terms of the action $I_g$ as before\footnote{In addition
to the equation in footnote 5, eq.(44) also implies that
$\sqrt{\mathop{g}\limits^4}(R^2)^{\prime \prime}/R^2 =
\partial_5[\sqrt{\mathop{g}\limits^4}R^{\prime}/R]$.}. Thus,
$A_0$ and $\xi$ are not fixed by field equations in this general
case either.

 \vspace{10pt}
\centerline {\Large \bf  Summary }

\vspace{10pt} We have examined in the category of 5d Kaluza-Klein
theory a gauge equivalence of $x^5$-dependent solutions each of
which describes in the 4d sector an arbitrarily evolving
isotropic, homogeneous universe with a zero, or non-zero pure
gauge. The main points we have observed are: (1)Within a certain
time scale $\tau_c$ (which is characterized by the
compactification radius $R_c$) any arbitrarily evolving 4d
universe with an appropriate pure gauge is allowed to exist by
field equations and these isotropic, homogeneous universes are all
gauge equivalent as long as they are of the same topology. (2)In
this case the pure gauge $A_0$ plays a role of the 4d effective
matter source as does the dynamics of the fifth
dimension\footnote{It is well-known that the dynamics of the fifth
dimension acts as a 4d effective radiation source. See, for
instance, ref.\cite{555}.}. A pure gauge has its own energy
density which manifests itself when the metric components have
$x^5$-dependence. (3)Outside $\tau_c$ the evolution of the
universe is set by field equations. In particular, for $T^{AB}=0$
the only allowed state of universe is a radiation-dominated
universe with an arbitrary pure gauge which can be removed by a
coordinate transformation.

The above result naturally leads us to a certain conjecture which
perhaps makes our point more clear. The suggested conjecture is
that it may be totally meaningless in 5d Kaluza-Klein theory (or
even in any higher-dimensional theory) to distinguish one 4d
universe from another (as long as they are of the same topology)
within a certain time region $\tau_c$ which is expected to be of
order of the compactification scale, because there is no known
gauge-fixing mechanism to select a preferred universe among the
infinite number of gauge equivalent universes. The physics in such
a region is always subtle and complicated due to the existence of
such as an initial singularity, or a quantum fluctuation etc. The
above conjecture may provide a possibility of avoiding such
difficulties. For instance, we can avoid the initial singularity
by dealing with a gauge equivalent static universe with no initial
singularity, instead of dealing with an evolving universe with
initial singularity. This amounts to say that the initial
singularity is alleviated due to the existence of the internal
dimension(s) to the extent of its size. In short, the above
conjecture tells that within $\tau_c$ we do not need to consider
detailed dynamics or matter contents of 4d universes. We may
simply ignore them!

\vspace{15pt} \hspace{-0.6cm}{\Large \bf Acknowledgments}

\vspace{10pt} This research was supported by the Kyungsung
University research grants in 2002.

\end{document}